\documentclass[showpacs,twocolumn]{revtex4}
\usepackage{amssymb}
\usepackage{amsmath}
\usepackage{amstext}
\newcommand*\de{\mathrm{d}}
\newcommand*\De{\mathrm{D}}
\renewcommand*\epsilon{\varepsilon}
\renewcommand*\phi{\varphi}
\renewcommand*\theta{\vartheta}

\begin{document}
  
\title{Intrinsic momentum in Poincar\'e gauge theory} 

\author{M. Leclerc} 
\affiliation{Section of Astrophysics and Astronomy, 
Department of Physics, \\ University of Athens, Greece}  
\begin{abstract}
While it is generally accepted, in the framework of Poincar\'e 
gauge theory, that the Lorentz connection couples minimally to  
spinor fields, there is no  general agreement on 
the coupling of the translational gauge field to fermions. We will show that 
the assumption that spinors carry a full Poincar\'e representation leads to 
inconsistencies, whose origins will  be traced back by 
considering the Poincar\'e group both as the contraction  of the de Sitter 
group, and as a subgroup of the conformal group. As a result, the
translational fields do not minimally couple to fermions, and consequently, 
fermions do not possess an intrinsic momentum. 
\end{abstract}
\pacs{04.50.+h, 11.15.-q} 
\maketitle

\section{Introduction}

In Poincar\'e gauge theory, and especially in the more widely known 
Einstein-Cartan theory, a majority of authors follows the standard 
references \cite{hehlA, hehlB}, considering scalar and gauge fields 
(other than the gravitational  fields) 
as invariant under the Poincar\'e gauge group, while  spinor fields 
are supposed to carry a Lorentz representation. This leads to a consistent 
theory, and one could  justify the invariance under translations by the fact 
that the translational symmetry  will be  broken anyway,  and that therefore 
 the Lorentz group is  the only physical gauge group.

On the other hand, it is certainly not unnatural 
 to consider fermions carrying  a full 
Poincar\'e representation. As a result, the covariant derivative of spinor
 fields will contain
the full Poincar\'e connection, leading therefore  to a minimal coupling 
to the translational gauge fields. Once  the translational gauge freedom 
is broken, the fermions   reduce to the usual Lorentz spinors, but 
the minimally coupled  translational field (which becomes a simple 
Lorentz vector-valued one-form) remains in the Lagrangian. Both approaches 
are therefore not equivalent.  

The  first attempt to treat gravity as gauge theory goes back to Utiyama 
\cite{utiyama}, with the gauge group taken to be the Lorentz group, while 
the tetrad fields had to be introduced ad hoc.  
The precise relation between tetrad fields and the translational gauge fields 
was clarified much later, which could explain why the possibility of
fermions carrying a Poincar\'e representation is not even mentioned in 
 most of the literature (see,  e.g.,  
\cite{hehlA, tseytlin, hehlB}). Another good 
reason for this can also be seen from the experimental side. While it is 
a well established fact that fermions possess, apart from their orbital 
angular momentum, also an intrinsic spin  momentum, their is no evidence in 
favor of a concept like \textit{intrinsic momentum}.

Nevertheless, the alternative approach, with spinors carrying in addition a 
representation of the translational part of the Poincar\'e group, 
can be found in \cite{grignani} and,  
more recently, in \cite{tiembloA, tiembloB}. 

In this paper, we will conclude that the coupling of the 
translational gauge fields to fermions faces problems, and that only 
 the classical approach with Lorentz spinors is free of inconsistencies. 
This conclusion is further supported by the following. 
If the Poincar\'e gauge theory is treated as a subcase of the theory based 
on the conformal group $SO(4,2)$,  it turns out that, 
on one hand, in the Lorentz
invariant groundstate, no minimal coupling of the translational gauge 
field occurs, and moreover,  no corresponding Poincar\'e invariant 
groundstate exists. 
In other words, the residual Lorentz theory cannot be 
interpreted as the result of a symmetry breakdown of the translational 
gauge freedom in the framework of a Poincar\'e gauge theory. 

On the other hand, if we start from the de Sitter group $SO(4,1)$, and 
take the limit to the Poincar\'e group via a Wigner-In\"on\"u contraction, 
we end up with a consistent Poincar\'e 
invariant theory, with, however, no direct coupling of the translational 
gauge fields to the spinor fields. 

The article is organized as follows. In the next section, we briefly review 
the basic concepts of Poincar\'e gauge theory, focusing mainly 
on the translations and their relation to the tetrad fields. Then, in 
section III, we point out the inconsistencies arising from the minimal 
coupling of the translational field to fermions, and finally, in sections 
IV and V, we treat Poincar\'e theory as a limiting case of the 
de Sitter and the conformal gauge theories, respectively. 

\section{Translations in Poincar\'e gauge theory}

We start with a Poincar\'e connection one-form 
$(\Gamma^{ab}, \Gamma^a)$ (where $a=0,1,2,3,\ \Gamma^{ab} = - \Gamma^{ba}$)
which transforms under an infinitesimal Poincar\'e transformation with 
coefficients $(\epsilon^{ab}, \epsilon^a)$ as 
\begin{equation}
\delta \Gamma^{ab} = - \De \epsilon^{ab}, \ \ 
\delta \Gamma^a =  - \De \epsilon^a + \epsilon^a_{\ b} \Gamma^b.
\end{equation}
As is customary, we  use the symbol $\De$ to denote the Lorentz 
covariant derivative, i.e., $\De  \epsilon^{ab} = \de \epsilon^{ab}
+ \Gamma^a_{\ c}  \epsilon^{cb} + \Gamma^b_{\ c}  \epsilon^{ac}$.
It turns out that with those gauge fields alone, it is not possible to 
construct a consistent Lagrangian 
(see \cite{hehlB, grignani, mielke, leclerc}) and the introduction of an 
additional Poincar\'e vector $y^a$, transforming as 
\begin{equation}
\delta y^a = \epsilon^a_{\ b} y^b + \epsilon^a,  
\end{equation}
is necessary. Then, one defines the tetrad field 
\begin{equation}
e^a = \Gamma^a + \De y^a, 
\end{equation} 
which is invariant under translations ($\delta e^a = \epsilon^a_{\ b} e^b$). 
In order to avoid the introduction of an arbitrary  
length parameter, we suppose that $\Gamma^a$ is dimensionless (as 
opposed to $\Gamma^{ab}$ which has dimensions $L^{-1}$), while 
$\epsilon^a $ and $y^a$ have dimension $L$.  
The  gravitational Lagrangian is now constructed from $e^a$, as well as 
from the curvature 
$R^{ab} = \de \Gamma^{ab} + \Gamma^a_{\ c} \wedge \Gamma^{cb}$ and 
the torsion $T^a = \de e^a + \Gamma^a_{\ b}\wedge  e^b$. 

The origin of the so-called Poincar\'e coordinates $y^a$ has been traced back  
in  \cite{tseytlin,mielke, tres} to the non-linear realization of the 
translational part of the gauge group. There is a close connection 
between non-linear realizations and the Higgs symmetry breaking mechanism
(see \cite{tseytlin}). Indeed, it is possible to treat $y^a$ as a Higgs 
field, with groundstate constitution $y^a = 0$. This groundstate  
 breaks the translational symmetry and leaves us with a residual 
Lorentz symmetry. The details of such an approach have been elaborated 
in \cite{leclerc}. 

A Lorentz spinor transforms under a Poincar\'e transformation as 
\begin{equation}
\psi \rightarrow e^{(-i/4) \epsilon^{ab} \sigma_{ab}}\psi, 
\end{equation}
while a Poincar\'e spinor would transform as 
\begin{equation}
\psi \rightarrow e^{(-i/4) (\epsilon^{ab} \sigma_{ab}+ \epsilon^a P_a) }\psi, 
\end{equation}
where $(\sigma_{ab}, P_a)$ are the generators of the Poincar\'e group. 
In order to be 
compatible with our dimension conventions $[\epsilon^a] = [L]$, we have 
to require $[P_a] = [L^{-1}]$.  

Especially, in the classical approach with Lorentz spinors, the generators 
are taken to be  $\sigma_{ab} = 
\frac{i}{2} [\gamma_a, \gamma_b]$, and the following Dirac Lagrangian 
\begin{equation}
\mathcal L = -\frac{i}{12} \epsilon_{abcd} e^a \wedge e^b \wedge e^c \wedge
(\bar \psi \gamma^d \De \psi - \De \bar \psi \gamma^d \psi), 
\end{equation}
where $\De \psi = \de \psi - (i/4) \Gamma^{ab} \sigma_{ab} $ is easily shown 
to be Poincar\'e invariant. 
We omit the 
mass term, which is unrelated to  our discussion. 

It is clear that every Lagrangian constructed only out  of 
$e^a, \Gamma^{ab} $, invariants (like the Maxwell field $A$ or scalar fields)
as well as  Lorentz spinors, will trivially be invariant under translations. 
Moreover, in view of the relation (3), no additional field equation will 
arise from variation with respect to $y^a$. The resulting equation 
will simply be the covariant derivative of the equation resulting from the 
$\Gamma^a$ variation and is thus identically satisfied. 
 Further, instead of varying with respect to $\Gamma^a$, 
one can equivalently vary with respect to $e^a$. As a result, one can 
simply forget about $y^a$ and identify directly $\Gamma^a $ with $e^a$, 
resulting in an effective Lorentz gauge theory. Thus, in some  
sense, the gauging 
of the translations and the introduction of the Higgs field $y^a$ is 
basically  a way to explain the presence of the tetrad field, but for the 
rest, 
one considers only Lorentz transformations, just as Utiyama \cite{utiyama} 
 did right from the start. 

Things changes, however, as soon as we introduce a Poincar\'e spinor (5). 
Then, clearly, the translational invariance of a Lagrangian is not 
a priori guaranteed, and moreover, the use of a Poincar\'e covariant 
derivative will break the equivalence of the $\Gamma^a$ and the $e^a$ 
variations. Let us introduce, for convenience, the following 
notations: 
\begin{equation}
\mathcal P 
= e^{(-i/4) (\epsilon^{ab} \sigma_{ab}+ \epsilon^a P_a) },\ \ 
 \mathit \Lambda  = e^{(-i/4) \epsilon^{ab} \sigma_{ab} }.   
\end{equation}
Thus, we have for the Poincar\'e spinor $\psi \rightarrow \mathcal P \psi$. 
The covariant derivative $\nabla \psi $ transforming in the same way,  
$\nabla \psi \rightarrow \mathcal P \nabla \psi$, is given by  
\begin{equation}
\nabla \psi = \de \psi - \frac{i}{4} 
(\Gamma^{ab} \sigma_{ab} + \Gamma^a P_a)  
= \De \psi - \frac{i}{4} \Gamma^a P_a.  
\end{equation}
 We see that, if we have such a covariant derivative in our Lagrangian, 
the fields $\Gamma^a $ and $y^a$ do not only occur in the combination 
(3), and thus, the variation has to be carried out with respect to all 
independent fields $(\Gamma^{ab}, \Gamma^a, y^a, \psi )$. Together with 
 the additional 
field equation, we get a new conservation law (because the translational 
gauge invariance is not trivially satisfied anymore), leading to the 
so-called intrinsic momentum conservation. 

Such Lagrangians will be investigated in the next section.

\section{Dirac fields with intrinsic momentum}

Recently, in \cite{tiembloA, tiembloB}, the following Dirac Lagrangian 
has been proposed 
\begin{equation}
\mathcal L = -\frac{i}{12} \epsilon_{abcd} e^a \wedge e^b \wedge e^c \wedge
(\bar \psi \gamma^d \nabla \psi - \nabla \bar \psi \gamma^d \psi), 
\end{equation}
where 
\begin{equation}
\nabla \psi = \de \psi - \frac{i}{4} 
(\Gamma^{ab} \sigma_{ab} + e^a P_a)  
= \De \psi - \frac{i}{4} e^a P_a,   
\end{equation}
and the Poincar\'e generators are taken to be 
\begin{equation}
\sigma_{ab} = \frac{i}{2} [\gamma_a, \gamma_b], \ \ 
P_a = m \gamma_a (1+ \gamma_5). 
\end{equation}
This Lagrangian is supposed to be invariant under the non-linear 
realization of the Poincar\'e group with Lorentz structure group. 
In other words, it is invariant after the reduction of the symmetry 
group to the Lorentz group (see \cite{mielke, tres, leclerc} for 
the concept of non-linear realizations in the context of Poincar\'e gauge
theory). That (9) is indeed Lorentz invariant is not hard to show, 
since the only difference to (6) comes from the coupling term 
$\sim \bar \psi e^a P_a \psi $  contained in 
the covariant derivatives, leading to a 
mass term for the Dirac particle (see \cite{tiembloA}).  

However, it is usually understood that different choices of the stability 
subgroup in the non-realization scheme should lead to physically 
equivalent results, since only the initial symmetry group 
(in our case the Poincar\'e group) is physically relevant. For instance, in 
\cite{tiembloB}, the same Lagrangian (9) is also considered with  
stability group $SO(3)$ instead of the Lorentz group. Especially, one 
could also consider the stability subgroup to be the Poincar\'e group 
itself. In the more physical approach, using the Higgs fields $y^a$, 
this simply means that the Lagrangian should be independent of any 
specific gauge  we implement on $y^a$. This holds independently  
of how the specific gauge arises, be it simply a convenient 
choice, or be it the 
result of a symmetry breaking groundstate, like $y^a = 0$. In any case, 
before we can choose a gauge, or before a gauge symmetry can break down, 
there has to be an initial Lagrangian invariant under the whole gauge 
group. We cannot argue that the translational symmetry is always 
broken, and that therefore, it is enough to be in the possession of the 
residual, Lorentz invariant Lagrangian.    

After the above considerations, it is clear that (9)  corresponds to 
the Lagrangian in the gauge $y^a = 0$ (i.e., the gauge where 
$\Gamma^a  = e^a$). It remains to find the gauge invariant Lagrangian, 
reducing to (9) in the specific gauge. It is clear that the covariant 
derivative (10) has to be replaced by (8). However, it is easily shown 
that (9) will still not be invariant under translations. 

This may seem quite strange, since the usual gauge prescription tells 
us to couple the gauge field minimally to $\psi$, as in (10), and 
that the Lagrangian will then be invariant if and only if the gauge fields 
transform under the adjoint representation of the group (which is also 
the case, see (1)). So, what did we do wrong? The answer is found in 
a fundamental difference of gravitational gauge theories and conventional 
Yang-Mills theory: The transformation (5), and also (4), acts directly 
in Dirac space, while conventional gauge transformations 
 take place  in an additional, 
inner space. Why does that lead to a problem? Well, apart from 
the field $\psi$, there are other objects in (9) that are defined in 
Dirac space, namely the $\gamma$-matrices. We will thus have to 
discuss their transformation behavior. 

Actually, the same problem is also present in the Lagrangian (6), but 
it turns out that, under Lorentz transformations, 
\begin{equation}
\mathit \Lambda \gamma^a \mathit \Lambda^{-1} \Lambda^a_{\ b} = \gamma^a, 
\end{equation}
where infinitesimally, $\Lambda^a_{\ b} = \delta^a_b + \epsilon^a_{\ b}$. 
This allows us to consider $\gamma^a$ as invariant, or simply as 
constant matrices. As a result, the Lorentz generators $\sigma_{ab}$ 
too will be invariant. This is a general feature of gauge theories. 
Namely, if $\psi $ transforms as 
$\psi \rightarrow \exp(i \epsilon^{\alpha}\sigma_{\alpha})\psi$, where
$\sigma_{\alpha}$ are the generators of the gauge group, then, the 
generators themselves, transforming as $\sigma_{\alpha} \rightarrow 
\exp(i \epsilon^{\gamma}\sigma_{\gamma})\sigma_{\beta} \exp(-i\epsilon^{\gamma}
\sigma_{\gamma})G^{\beta}_{\ \alpha}$, will be invariant if and only if 
$G^{\alpha}_{\ \beta} = \delta^{\alpha}_{\beta} - \epsilon^{\gamma}
\lambda_{\gamma \beta}^{\ \ \alpha}$, where 
$\lambda_{\gamma \beta}^{\ \ \alpha}$ are the structure constants of the
group. This means simply that, with respect to the index $\alpha$, 
$\sigma_{\alpha}$ transforms under the adjoint representation (just as 
 the gauge field $A^{\alpha}$, apart form the inhomogeneous  term
$\de \epsilon^{\alpha}$). Note that in the Lorentz case, each index $\alpha$ 
corresponds to an antisymmetric  pair $(ab)$. Similar, one can find a 
double indexed adjoint representation of the Poincar\'e group, starting with  
a 5d matrix form vector representation (see, e.g.,  \cite{leclerc}). 

Thus, formally, we can be sure that the generators $\sigma^{A}$, where 
$A$ labels the ten Poincar\'e generators, are invariant (as they have 
too), if they transform as $\sigma^A \rightarrow 
\mathcal P \sigma^B \mathcal P^{-1} 
P^A_{\ B}$, where $P^A_{\ B}$ is the adjoint 
representation of the Poincar\'e  group. This ensures, for instance, 
that $\sigma_{ab}$ is invariant, but still does not allow us to conclude 
on $\gamma^a$!

In view of the difficulties with the double indexed quantities, it is 
convenient to grab the problem from the other side, namely from (9). 
It is immediately clear that (9), with the covariant derivative (8), 
can only be Poincar\'e invariant if $\gamma^a$ transforms as follows
\begin{equation}
\gamma^a \rightarrow \mathcal P \gamma^a \mathcal P^{-1} \Lambda^b_{\ a}, 
\end{equation}
where $\Lambda^a_{\ b} $ is the usual vector representation of the 
Lorentz group, $\Lambda^a_{\ b} = \delta^a_{\ b} + \epsilon^a_{\ b}$.      
This is simply because $\psi$ and $\nabla \psi$ transform under the 
fundamental representation of the Poincar\'e group, $\bar \psi $ and 
$ \nabla \bar \psi$  under its  inverse, and $\epsilon_{abcd}e^a e^b e^c$ 
as Lorentz vector (see (3)). 

However, it can also be seen, using the explicit generators (11), 
 that under (13), if $\mathcal P$ contains a 
translation, $\gamma^a$ is not invariant. Moreover, neither is 
$[\gamma^a, \gamma^b]$ invariant. From this result, several  problems arise. 
First, the generators (11), which have to be invariant, cannot be 
defined with the same $\gamma$-matrices that appear in the Lagrangian and 
are gauge dependent.  
This leads to a second problem, namely we will  not be able 
to determine the commutation relations between $\gamma^a$ and the 
generators. Finally, only in specific gauges will $\gamma^a$ be 
constant. In general, $\gamma^a$ will depend on the spacetime coordinates, 
and consequently will have to be treated as dynamical field. This does 
certainly not look like a consistent theory. 

We conclude that (9), with covariant derivative  (8),  
is not the Poincar\'e invariant form 
of (9) with derivative (10). An alternative attempt can be found in 
\cite{grignani}. They use the Lagrangian (9), again with the Poincar\'e 
covariant derivative (8), but replace $\gamma^a$ by 
\begin{eqnarray}
\tilde \gamma^a = \gamma^a + (i/4)m y^b[\gamma^a \gamma_b(1+\gamma_5)- 
\gamma_b \gamma^a(1- \gamma_5)] \nonumber \\ +  (m^2/4) (y^a \gamma_b y^b - 
\frac{1}{2} |y^2| \gamma^a)(1+ \gamma_5). 
\end{eqnarray}
In the gauge $y^a = 0$, this reduces  to $\gamma^a$, and thus (14) too 
is a covariant generalization of (9).  Unfortunately, the authors do 
not tell us how $\gamma^a $ and $\gamma^5 $ are supposed to transform 
under gauge transformations, 
but they claim that the Lagrangian is invariant under 
the full Poincar\'e group. 

It is again straightforward to show that for this to be the case, $\tilde
\gamma^a$ has to transform as 
\begin{equation}
\tilde \gamma^a \rightarrow \mathcal P \tilde 
\gamma^a \mathcal P^{-1} \Lambda^b_{\ a}.  
\end{equation}
We were not able to find the transformation behavior for $\gamma^a$ that 
leads to (15) for $\tilde \gamma^a$. One can, however, easily verify 
that claiming  $\gamma^a $ to be invariant, does not 
lead to the above result. On the other hand, the generators (11) have to be 
invariant. Therefore, it would rather be a surprise if someone can come up 
with a non-trivial transformation behavior for $\gamma^a$ leading both to 
(15) and to invariant generators. (Especially, 
it would be strange if the authors of \cite{grignani}  had known of 
such a transformation and did not find it necessary to write it down.) 
However, even if such a transformation exists, the fact that $\gamma^a$ is 
not invariant (and will thus be coordinate dependent in certain gauges), is 
enough for the theory to be inconsistent, as outlined above. 
 
We have to conclude that no consistent theory with translational 
gauge fields coupling to the Dirac field has yet been found. 
This does not mean that it is impossible to construct such 
a theory, but it shows at least that the straightforward approach 
does not lead to the expected success. 

\section{The de Sitter group} 

Part of the difficulties in  constructing  a consistent Poincar\'e 
invariant theory comes from the non semisimple nature of the group.
It might therefore be instructive to start from rotational groups 
like $SO(4,1)$ or $SO(4,2)$ and consider the Poincar\'e theory as 
a limiting case of those theories. 

The most promising candidate is probably the de Sitter group, since its
algebra  
has the same dimension as that of the Poincar\'e group. The Poincar\'e 
group is recovered through the application of the so-called Wigner-In\"on\"u 
contraction. The de Sitter gauge theory has been discussed in full detail 
in  \cite{stelle}. The article contains a discussion of the non-linear 
realization scheme as well as of the possibility of a Higgs 
symmetry breaking mechanism. 
The latter has been further investigated in \cite{leclerc}. 

In this section, capital indices $A,B\dots$ take the values $0,1,2,3,5$, 
while the four dimensional part is denoted by $a,b\dots$ as before, i.e., 
 $A = (a, 5)$. We start with a de Sitter connection  $\Gamma^{AB}$, 
transforming as 
\begin{equation}
\delta \Gamma^{AB}  = - \nabla \epsilon^{AB}, 
\end{equation}
where $\nabla $ denotes the de Sitter invariant derivative. The de 
Sitter transformations (of the infinitesimal form 
$G^A_{\ B} = \delta^A_{\ B} + \epsilon^A_{\ B}$) leave the de Sitter 
metric $\eta_{AB} = diag(+1,-1,-1,-1,-1)$ invariant. A de Sitter spinor 
transforms as 
\begin{equation}
\psi \rightarrow e^{(-i/4) \epsilon^{AB} \sigma_{AB}}\psi, 
\end{equation}
where the generators are taken to be $\sigma_{AB} = (i/2)[\gamma_A,
\gamma_B]$, with $\gamma_A = (\gamma_a, i \gamma_5)$, where $\gamma_5 = 
-(i/4!) \epsilon_{abcd}\gamma^a \gamma^b \gamma^c \gamma^d$.  
 As in the Poincar\'e case, a Higgs field $y^A$ is introduced, 
transforming as 
\begin{equation}
\delta y^A = \epsilon^A_{\ B} y^B. 
\end{equation}   
We suppose that the complete theory contains a Higgs sector leading to 
the groundstate $y^A = (0,0,0,0,l)$, invariant under the Lorentz group 
only (see \cite{leclerc}). 
The parameter  $l$ is an explicit ingredient of the theory. It has the 
dimensions of a length and in the limit $l \to \infty $, the theory will 
reduce to a Poincar\'e gauge theory. 

Let us introduce the following one-forms
\begin{equation}
E^A = \nabla y^A, 
\end{equation}
which reduce, in the groundstate, to $E^A = (l \Gamma^a_{\ 5}, 0)$. 
In view of this, and its homogeneous transformation behavior under the 
residual Lorentz group, we identify  $l \Gamma^a_{\ 5} $ with the 
tetrad $e^a$. 

A gauge invariant Dirac-type Lagrangian for the de Sitter spinor is 
readily written down 
\begin{equation}
\mathcal L =\!\! - \frac{i}{12} \epsilon_{ABCDE}E^A\!\wedge E^B\! \wedge E^C 
\!\wedge ( \bar \psi \gamma^D \nabla \psi - \nabla \bar \psi \gamma^D \psi) 
y^E\!/l.
\end{equation}
The spinor derivative is defined as usual, by $\nabla \psi = \de \psi 
- (i/4) \Gamma^{AB} \sigma_{AB}$. In the gauge $y^A = (0, l)$, or, if 
you prefer, in the groundstate, $\mathcal L$ takes the form (9), 
with the derivative 
\begin{eqnarray}
\nabla \psi &=& \de \psi - (i/4) \Gamma^{AB} \sigma_{AB} \nonumber \\
&=& \de \psi - (i/4) \Gamma^{ab} \sigma_{ab} + l^{-1}(i/2) 
e^a \sigma_{a5} \nonumber 
\\ &=&  \De \psi  - l^{-1}(i/2) e^a \gamma_a
\gamma_5 \nonumber 
\end{eqnarray}
The Lagrangian therefore takes the form 
\begin{eqnarray}
\mathcal L = -\frac{i}{12} \epsilon_{abcd} e^a \wedge e^b \wedge e^c \wedge
(\bar \psi \gamma^d \De \psi - \De \bar \psi \gamma^d \psi)\nonumber \\
+ \frac{1}{12} \epsilon_{abcd} e^a\! \wedge e^b\! \wedge e^c\! \wedge (e^e/2l) 
(\bar \psi (\gamma^d \gamma_e\! -\! \gamma_e \gamma^d) \gamma_5)\psi, 
\end{eqnarray} 
where $\De $ the Lorentz covariant derivative. Clearly, the terms in the 
second line 
vanish, and thus, the Lagrangian reduces to the classical, Lorentz 
covariant Lagrangian (6), without a direct coupling of $e^a$ to the 
spinor field, in contrast to the attempt (9). 

What does that mean? Well, even without taking  the Poincar\'e limit 
$l \to \infty$, in the Lorentz invariant groundstate, there is no direct 
coupling of the pseudo-translational part $\Gamma^a_{\ 5} = l^{-1} e^a$ of 
the connection to the spinor field. Therefore, if we take the 
limit $l \to \infty$  in the de Sitter 
invariant form  (20), this will still  lead to a theory which, in the Lorentz 
invariant groundstate, will not present a minimal coupling of the 
translational fields to the spinor. In other words, whatever the coupling 
of the translational field looks like in the initial Poincar\'e Lagrangian, no 
coupling of the form (10) will remain in the Lorentz invariant gauge. 
But we know already the Poincar\'e invariant form of such a theory, it 
is the Lagrangian (6) itself, if we assume simply that $\psi $ carries 
only a Lorentz representation, as we have argued in section II. Indeed, 
the same Lagrangian emerges by taking the limit $l \to \infty $ in (20). 
This involves, however, a careful parameterization of the group   
generators and the fields in terms of  $l$, in a way that the 
corresponding Poincar\'e structure results in the limit $l \to \infty$. 
The procedure is standard (see \cite{inonu}), and is not 
needed for our purpose. 

Summarizing, treating the Poincar\'e group as Wigner-In\"on\"u contraction 
of the de Sitter group clearly suggests that in Poincar\'e gauge theory, 
the spinor fields do not carry a representation of the full group, 
but only of the Lorentz part, and consequently, couple minimally only 
to the Lorentz connection. In short, they do not possess intrinsic momentum. 

Our result was achieved by analyzing the gauge theory of the 
group $SO(4,1)$. As far as the limit to the Poincar\'e group 
is concerned, one could expect that the $SO(3,2)$ theory leads to 
identical results. 
However, there are interesting differences between both theories. 
One might think that the $SO(3,2)$ spinor representation 
 emerges from the $SO(4,1)$ representation simply through the 
replacement $\gamma_5 \rightarrow i \gamma_5$, changing in this way 
the signature in the metric $2 \eta_{AB} = \gamma_A \gamma_B+ \gamma_B 
\gamma_A$. It turns out that this is not correct for the following 
reason. In order to define the conjugate spinor $\bar \psi$ 
transforming with the inverse $G^{-1}$ of the gauge group, 
we write $\bar \psi = \psi^{\dagger} \gamma $. Then, $\gamma$ 
has to transform as $\gamma \rightarrow (G^{\dagger})^{-1}\gamma G^{-1}$. 
Since we do not want $\gamma$ to be a dynamical field, it should be 
invariant. Thus, infinitesimally, for $G = \exp{(i \sigma)}$, we 
 require  
\begin{equation}
 \sigma^{\dagger} \gamma = \gamma \sigma.  
\end{equation}
The only hermitian matrices, up to a global factor,   
satisfying this relation for all the Lorentz generators 
$\sigma_{ab}$ are $\gamma = \gamma_0 $ and $\gamma = I \gamma_5 \gamma_0 $.

For the usual choice $\gamma = \gamma_0$,  
 is is easily verified that the only 
transformations $G = \exp{(i \sigma)}$ satisfying (22) 
are given by {\bf real} linear combinations  of 
\begin{equation}
\sigma = (\sigma_{ab}, \gamma_a, \gamma_a \gamma_5, i \gamma_5)
\end{equation}
The largest possible gauge group with a four dimensional representation 
is therefore the conformal group. The generators 
$(\sigma_{ab}, \gamma_a \gamma_5)$ span the algebra of the de Sitter 
group $SO(4,1)$, 
while $(\sigma_{ab}, \gamma_a)$ span the anti-de Sitter algebra $SO(3,2)$. 
However,  that  the latter, as opposed to case $SO(4,1)$, 
is not a spinor representation in the strict sense, i.e.,  
 we cannot write $ 2\sigma_{AB} = i [\gamma_A, \gamma_B] $ and $2 \eta_{AB} = 
 \{\gamma_A, \gamma_B\}$ for some $\gamma_A$. This, however, has the 
consequence there is no set of matrices such that  $\gamma^A$ 
is  a vector, i.e, we do not have 
the relations $[\gamma^C, \sigma_{AB}] = 2 i (\delta^C_A \gamma_B-
\delta^C_B \gamma_A)$ (meaning that  $\gamma^A$ 
is invariant under gauge transformations, as in (12)).  

Thus, on one hand, the generators $(\sigma_{ab}, i \gamma_a \gamma_5)$  
 span the algebra of the group $SO(3,2)$, but  do not leave invariant 
the spinor metric $\gamma_0$. On the other hand, 
the  generators 
 $(\sigma_{ab}, \gamma_a) $ span the same algebra, leave the metric 
invariant, but do not allow for the construction of a Lagrangian 
in the form (20), because we do not have an invariant set of 
$\gamma$-matrices.

On the other hand,  
if we use instead $\gamma = i \gamma_5 \gamma_0 $ as spinor metric, 
then the transformations allowd by (22) are generated by 
 $\sigma = (\sigma_{ab}, i \gamma_a, i \gamma_a \gamma_5, i \gamma_5)$. 
Thus, in that case, the generators $(\sigma_{ab}, i \gamma_a \gamma_5)$  
give rise to a true spinor representation of $SO(3,2)$, but now, 
the construction of an $SO(4,1)$ invariant Lagrangian is not possible. 
Both groups are thus simply interchanged and the conclusions concerning 
the translations are identical.

\section{The conformal group}

The conformal group  contains the Poincar\'e group as a subgroup, and 
could  thus reveal an alternative approach to the coupling of the 
translational part of the connection. In this section, capital indices 
take values $A= (0,1,2,3,5,6)=  (a,5,6)$. The conformal group $SO(4,2)$ 
is defined to leave the metric $\eta_{AB}= diag(+1,-1,-1,-1,-1,+1)$ 
invariant. 

Although there exists a four dimensional representation of the 
conformal group, with the algebra spanned, e.g., 
 by the generators (23), 
we face again the problem that there is no corresponding set of 
invariant $\gamma$-matrices for the construction of the Lagrangian. 
As outlined in the previous section, the largest possible 
transformation group is either given by $SO(3,2)$ or by $SO(4,1)$, depending 
on the choice of the spinor metric. 

Therefore, we necessarily have to consider the eight dimensional 
spinor representation of $SO(4,2)$. The corresponding Dirac algebra 
can be found, e.g., in \cite{mack}. We define the 6 matrices $\beta^A$ by 
\begin{equation}
\beta^a = \gamma^a \sigma_3,\ \ \beta^5 = i \sigma_1, \ \ \beta^6 = \sigma_2,
\end{equation}
where $\gamma^a $ and $\sigma_i$ (i = 1,2,3) are defined as 
\begin{eqnarray*}
\gamma^a = \begin{pmatrix} \gamma^a & 0 \\ 0 & \gamma^a\end{pmatrix},\ \  
 \sigma_1 = \begin{pmatrix} 0&1\\1&0 \end{pmatrix},\\ 
 \sigma_2 = \begin{pmatrix} 0&-i\\i&0 \end{pmatrix},\ \  
 \sigma_3 = \begin{pmatrix} 1&0\\0&-1 \end{pmatrix},
\end{eqnarray*}
composed of $4\times 4$ block matrices. The $SO(4,2)$ generators are now 
given by 
\begin{equation}
\sigma_{AB} = \frac{i}{2}[\gamma_A, \gamma_B].
\end{equation}
The generators 
$(\sigma_{ab}, P_a)$, with  $P_a = (1/2)(\sigma_{a5}+\sigma_{a6})$,  
 span the Poincar\'e subalgebra, which is what we are interested in. 

The 8 component spinor $\psi$ transforms formally as in (17). Next, 
define the spinor metric $\gamma = \gamma_0 \sigma_1$, i.e., the adjoint 
spinor is given by 
\begin{equation}
\tilde \psi = \psi^{\dagger} \gamma = \psi^{\dagger} \gamma_0 \sigma_1
= \bar \psi \sigma_1.  
\end{equation}
The relation (22) for $\gamma$ is easily verified for the fifteen 
generators (25), and thus, $\tilde \psi $ transforms under the inverse 
of the fundamental representation. (We reserve the notation $\bar \psi $ 
for $\psi^{\dagger} \gamma_0$, which will be useful later on.) 

The gravitational gauge theory of the conformal group, in the fashion 
of Stelle and West \cite{stelle}, has been elaborated  in
\cite{kerrick}. The connection transforms as in (16). 
In contrast to the Poincar\'e and de Sitter cases, we 
need two Higgs vectors $y^A$ and $z^A$ in order to break down the 
symmetry to the Lorentz group.  

Let us introduce the quantities 
\begin{eqnarray}
E^A\!\! &=&\!\! \frac{\sqrt{|ab|}}{2} \  
\nabla (z^A\!/b - y^A\!/a) \nonumber \\
&=&\!\!
\frac{\sqrt{|ab|}}{2} \left[
\de (z^A\!\!/b  \! - \!  y^A\!\!/a) \!+\! \Gamma^A_{\ C} 
(z^C\!\!/b - y^C\!\!/a)\right],   
\end{eqnarray}
where $a,b$ are two length parameters of the theory. It 
is possible, alternatively, to define $E^A $ simply as $\nabla y^A$, 
but we prefer the more symmetric approach using both $y^A$ and $z^A$. 

The conformally invariant Dirac-type Lagrangian is found in the form 
\begin{eqnarray}
\mathcal L &=& - \frac{1}{12} \epsilon_{ABCDEF}E^A \wedge E^B \wedge E^C 
\nonumber \\&&\wedge 
( \tilde \psi \beta^D \nabla \psi - \nabla \tilde \psi \beta^D \psi) 
(y^E/a) (z^F/b), 
\end{eqnarray}
which is a straight forward generalization of (6) and (20).   
In \cite{kerrick}, the Higgs fields were required, a priori, to satisfy
the relations 
\begin{equation}
y^Ay_A = -a^2,\  z^Az_A =  b^2,\  z^A y_A = 0. 
\end{equation}
This is to be interpreted as groundstate configuration (see \cite{leclerc}).
Let us first 
take a look at the Lorentz invariant groundstate, in order to see whether  
(28) reduces indeed to the Dirac Lagrangian. 

The Lorentz invariant groundstate, compatible with (29), is given by 
$y^A= (0,0,0,0,a,0)$ and $z^A = (0,0,0,0,0,b)$. In this gauge, (27) 
reduces to $E^A = (e^a, \sqrt{|ab|}\ A/2, 
- \sqrt{|ab|}\ A/2)$, where 
$e^a = \sqrt{|ab|}( \Gamma^{a5} + \Gamma^{a6})/2$, and $A = \Gamma^{56}$. 
Let us also introduce the notation $B^a = \sqrt{|ab|}(\Gamma^{a5} - 
\Gamma^{a6})/2$. 
(Note that if 
we choose, e.g., $z^A=(0,0,0,0,0,-b)$, 
then, from (27),  $E^a \sim B^a$. 
This is not a problem, since the corresponding generators, $Q_a= (\sigma_{a5}-
\sigma_{a6})/2$, span, together with $\sigma_{ab}$, 
 the Poincar\'e subalgebra too. Thus, the role of $(e^a , P_a) $ 
and $(B^a, Q_a)$ is  simply inverted.)  

The residual Lorentz invariant Lagrangian takes the form 
\begin{equation}
\mathcal L = -\frac{1}{12} \epsilon_{abcd} e^a \wedge e^b \wedge e^c \wedge
(\tilde  \psi  
 \beta^d \nabla \psi - \nabla \tilde \psi  \beta^d \psi), 
\end{equation}
with 
\begin{eqnarray}
\nabla \psi &=&\!\! \de \psi - (i/4) \Gamma^{ab} \sigma_{ab} 
\nonumber \\ \!&\!&\!\!\!- (i/4) [ \frac{4}{\sqrt{|ab|}} (e^a P_a  
\!+ \! B^a Q_a) + 2 A
\sigma_{56}] \psi,  
\end{eqnarray}

It is straightforward to verify that 
$e^a$ and $B^a$ transform homogeneously under the residual Lorentz group, 
whereas $\Gamma^{ab}$ is a Lorentz connection and $A$ an invariant. 

Thus, apart from the additional fields $B^a$ and $A$,  corresponding 
to the parts of the conformal group that do not belong to the 
Poincar\'e subgroup (i.e., generated by 
$\epsilon^{a5} - \epsilon^{a6}$ and $ \epsilon^{56}$), 
this Lagrangian is formally  identical to the  
one proposed in \cite{tiembloA}, 
namely (9), with derivative (10). The role of the 
 mass parameter $m$ is now played by 
the inverse length parameter $1/\sqrt{|ab|}$. Apart from this 
formal resemblance, we have to take into account that our particle 
interpretation is based on irreducible representations of the Lorentz group, 
and thus, in order to get an idea what particles are described by (30), 
we have to write the Lagrangian  
explicitely in terms of four component spinors. 
Introducing components $\psi = (\psi_1, \psi_2)$, we find from (31)
\begin{equation}
\nabla \begin{pmatrix} \psi_1 \\ \psi_2  \end{pmatrix}
= \begin{pmatrix} \De \psi_1 + \frac{i}{\sqrt{|ab|}} B^a \gamma_a \psi_2
- \frac{1}{2} A \psi_1 \\ \De \psi_2 - \frac{i}{\sqrt{|ab|}} 
e^a \gamma_a \psi_1 + \frac{1}{2} A \psi_2 \end{pmatrix}.  
\end{equation}
Using $\nabla \tilde \psi = (\nabla \psi)^{\dagger} \gamma_0 \sigma_1$, 
we get 
\begin{eqnarray*}
\nabla \tilde \psi = (\De \bar \psi_2,\   
\De \bar \psi_1)  \\ +
(\frac{i}{\sqrt{|ab|}}e^a 
\bar \psi_1 \gamma_a + \frac{1}{2}A \bar \psi_2, \  
-\frac{i}{\sqrt{|ab|}} 
B^a \bar \psi_2 \gamma_a - \frac{1}{2}A \bar \psi_1),
\end{eqnarray*}
with $\De $ denoting the Lorentz covariant derivative. 
Putting this into (30), it turns out that the diagonal terms all 
cancel out and we are left with 
\begin{eqnarray}
\mathcal L = -
\frac{1}{12} \epsilon_{abcd} e^a \wedge e^b \wedge e^c  \nonumber \\
\wedge \left[\bar  \psi_2  
 \gamma^d \De \psi_1 - \De \bar \psi_2  \gamma^d \psi_1 
-\bar  \psi_1  
 \gamma^d \De \psi_2 + \De \bar \psi_1  \gamma^d \psi_2 \right.
\nonumber \\ \ \  \ \left. - \bar \psi_2 \gamma^d A \psi_1 
- \bar \psi_1 \gamma^d A \psi_2\right].  
\end{eqnarray}
It is now easy to check that $\mathcal L$ is hermitian. This justifies 
the choice of the omitted factor $i$ in  (28), as compared to (20). 
For the special class of solutions $\psi_2 = - i \psi_1$, we recover 
exactly the conventional Dirac Lagrangian (6), coupling to the 
Lorentz connection only. More generally, (33) leads to two Dirac type 
equations for $\psi_1$ and $\psi_2$ with a minimal coupling to 
the Lorentz connection and to the field $A$. The latter coupling is identical 
to that of the Maxwell field, which justifies the identification made by 
Kerrick \cite{kerrick}. The addition of a  mass term 
$\sim m \tilde \psi \psi \sim m(\bar \psi_1 \psi_2 + \bar \psi_2 \psi_1)$ 
is not possible, because it leads ultimately to an imaginary mass in the 
field equations. The correct way  is to add a term $\sim m \tilde 
\psi \sigma_{AB}\psi\ y^A z^B/(ab)$ which is hermitian in view of (22) 
and reduces to $\sim i m(\bar \psi_1 \psi_2-\bar\psi_2 \psi_1)$ in 
the groundstate. 
 The Dirac limit is again obvious for $\psi_2 = - i \psi_1$, and 
in general, we get two Dirac particles with identical mass and opposite 
charge. 

In any case, we see, just as in the case of the de Sitter 
group, that no minimal coupling of the translational gauge field 
occurs  in the residual Lagrangian. Thus, 
we have to conclude that the coupling in the form 
 (9) and (10) cannot be obtained from the reduction of a conformally 
invariant Dirac equation. 

Apart from the fact that the Lorentz invariant groundstate does not 
lead to a coupling in the form (10), their is a more fundamental problem. 
Namely, the Lagrangian (28) does not possess a Poincar\'e invariant 
groundstate. Indeed,  
 under a Poincar\'e 
transformation $\epsilon^{AB}$, with $\epsilon^{a5}- \epsilon^{a6}=
 \epsilon^{56} = 0$, 
 the Higgs field $y^A$ transforms as $\delta y^A = 
\epsilon^A_{\ B} y^B$, thus, in particular 
\begin{eqnarray}
 \delta y^a &=& \epsilon^a_{\ b} y^b + \epsilon^a_{\ 5} y^5 
+ \epsilon^a_{\ 6} y^6 \nonumber \\
&=& \epsilon^a_{\ b} y^b - \epsilon^{a5}(y^5 - y^6). 
\end{eqnarray}
The only possible Poincar\'e invariant groundstate therefore has 
to satisfy  $y^a = 0$ and $y^5 = y^6$. 
The remaining conditions, $\delta y^5 = \delta y^6 = 0$ are then 
automatically satisfied. The same applies to $z^A$. 
Thus, the groundstate reads 
\begin{equation}
y^A = (0,0,0,0, a,a),\ \ \ z^A = (0,0,0,0,b,b),
\end{equation}
which satisfies $y_Ay^A = z_A z^A = z_A y^A = 0$, but not the 
constraints (29).  However, it is not 
hard to see that, with such a groundstate, the Lagrangian (28) vanishes 
identically. (The same holds for the free Lagrangian of the gravitational 
fields themselves, as presented in \cite{kerrick}.) 

As a result, the only Poincar\'e invariant groundstate does not lead to 
a consistent theory. The other way around, in order to get a consistent 
theory, the Higgs fields have to satisfy  the constraints (29). Those 
constraints, however, do not allow for a Poincar\'e invariant groundstate 
(apart from the trivial one, $y^A = z^A = 0$).

\section{Conclusions}

We have argued that the assumption that spinor fields carry a Poincar\'e 
representation, and therefore couple minimally to the translational 
gauge fields, leads to inconsistencies. Although Lorentz invariant 
Lagrangians with a minimal coupling of the tetrad field to the 
spinor have been written down in the past, those theories cannot 
be considered as the result of a symmetry breakdown of the 
translational gauge freedom in the framework of a Poincar\'e gauge 
theory, and thus, neither as the result of a non-linear realization 
of the translational subgroup. We must therefore conclude that spinors, 
in the framework of Poincar\'e gauge theory, do not possess intrinsic 
momentum.    

Apart from the  direct analysis, we base our conclusion on the fact that, 
on one hand,  no minimal coupling of translational fields arises from the 
theory obtained by applying a Wigner-In\"on\"u contraction to the de Sitter 
gauge theory, 
while on the other hand, the gauge theory of the 
conformal group  does not possess a Poincar\'e invariant groundstate.  

As it seems, the only field in Poincar\'e gauge theory that 
carries a representation of the complete group is the Higgs field $y^a$, 
coupling (in a certain sense, minimally) to the full Poincar\'e 
connection   via $e^a = 
\de y^a + \Gamma^a_{\ b} y^b + \Gamma^a$. 
This underlines the special role played by 
the translational gauge field,  serving ultimately in the construction 
of the spacetime metric,  and the particular structure of 
gravitational theories in general.

\section*{Acknowledgments}

This work has been supported by EPEAEK II in the framework of ``PYTHAGORAS 
II - SUPPORT OF RESEARCH GROUPS IN UNIVERSITIES'' (funding: 75\% ESF - 25\% 
National Funds).

\end{document}